\documentclass[usenatbib,onecolumn]{mn2e}
\usepackage{amsfonts}
\usepackage{amsmath}
\usepackage{graphicx}
\usepackage{natbib}

\def\apjl{Astrophys.\ J.\ Lett.}
\def\mnras{Mon.\ Not.\ R.\ Astron.\ Soc.}

\def\apj{Astrophys.\ J.}

\def\apjs{Astrophys.\ J. Supp.}

\def\prd{Phys.\ Rev.\ D}

\def\physrep{Phys. Rep.}
\title[Halo abundances in the $f_{nl}$ model]
      {Halo abundances in the $f_{nl}$ model}

\author[T. Y. Lam, R. K. Sheth]
 {Tsz Yan Lam\thanks{E-mail:tylam@sas.upenn.edu, shethrk@physics.upenn.edu}  \& Ravi K. Sheth\footnotemark[1]\footnotemark[1]\\
 Center for Particle Cosmology, University of Pennsylvania, 
 209 S. 33rd Street, Philadelphia, PA 19104, USA }

\newcommand{\bm}[1]{{\mbox{\boldmath $#1$}}}

\begin{document}
\pagerange{\pageref{firstpage}--\pageref{lastpage}}

\maketitle

\label{firstpage}

\begin{abstract}
We show how the excursion set moving barrier model for halo abundances 
may be generalized to the local non-Gaussian $f_{nl}$ model.  Our 
estimate assumes that the distribution of step sizes depends on $f_{nl}$, 
but that they are otherwise uncorrelated.  Our analysis is consistent 
with previous results for the case of a constant barrier, and 
highlights some implicit assumptions.  It also clarifies the basis 
of an approximate analytic solution to the moving barrier problem 
in the Gaussian case, and shows how it might be improved.  
\end{abstract}


\section{Introduction}
Detections of non-gaussianity can discriminate between different 
inflation models \cite[e.g.][]{jm03}.  The local $f_{nl}$ model, 
where the primordial perturbation potential is 
\begin{equation}
 \Phi = \phi + f_{nl}(\phi^2 - \langle \phi^2\rangle),
 \label{eqn:fnl}
\end{equation}
where $\phi$ is a Gaussian potential field and $f_{nl}$ is a scalar, 
has been the subject of much recent study
 \citep[e.g.,][and references therein]{bko08,kp08,st08}.  
Constraints on this model tend to be of two types -- from the CMB 
\citep{cmb5yr,hikageetal08,yw08,mhlm08} 
and from large scale structures in the Universe 
\citep{kst99,mvj00,ssz04,sk07,is07,fnlverde,dalaletal08,mv08,cvm08,at08,shshp08,mcdonald08,tkm08,slosar08,grossi08,kvj08,fnlvincent,lamshethfnl,grossinfm,vala09}. 

One of the fundamental quantities of interest in such studies is the 
abundance of virialized dark matter halos.  \cite{ps74} suggested that 
the abundance of collapsed virialized halos may be estimated from the 
statistics of the initial fluctuation field.  They used the assumption 
that halos form from a spherical collapse to argue that such objects 
started out as sufficiently overdense regions in the initial fluctuation 
field.  The excursion set approach of \cite{bcek91} allows one to 
estimate halo abundances in Gaussian theories; in this context, 
the spherical collapse model is associated a barrier of constant height.  
If the collapse is triaxial, then the barrier height is stochastic 
with a mean that is not constant \citep{smt01}.  Ignoring the 
stochasticity but including the changing of the barrier height allows 
the excursion set approach to provide a simple parametrization of the 
effects of triaxial collapse on halo abundances \citep{st02}.
The main goal of the present work is to show how to generalize the 
moving barrier formulae of \cite{st02} to the local non-Gaussian 
$f_{nl}$ model. 

Section~\ref{section:def} provides explicit expressions for the one- 
and two-point distribution of the overdensity in $f_{nl}$ models.  
These are used, in Section~\ref{section:mf}, to estimate how the mass 
function of virialized objects is modified when $f_{nl}\ne 0$.  
This section also clarifies earlier work on the Gaussian ($f_{nl}=0$) 
case.  A final section summarizes our results.
Two Appendices provide useful approximations and other technical 
details.

%%%%%%%%%%%%%%%%%%%%%%%%%%%%%%%%%%%%%%%%%%%%%%%%%%%%%%%%%%%%%%%%%%%%%%%

\section{The local non-gaussian model} \label{section:def}
We are interested in models where the primordial perturbation 
potential is given by equation~(\ref{eqn:fnl}).
We will use $P_{\phi}(k)$ to represent the power spectrum of $\phi$; 
in what follows we will set $P_{\phi}(k)= Ak^{n_s-4}$, where 
$n_s\approx 1$, and $A$ is a normalization constant that is fixed by 
requiring that the rms fluctuation in the associated non-Gaussian 
initial density field (which we will define shortly) has value $\sigma_8$.
The power spectrum and bispectrum of the $\Phi$ field are 
\begin{align}
 P_{\Phi}(k) & = P_{\phi}(k) + \frac{2f_{nl}^2}{(2\pi)^3}
   \int {\rm d}\, {\bm q} \left[P_{\phi}(q)P_{\phi}(|{\bm k} - {\bm q}|)
  - P_{\phi}(k)P_{\phi}(q) - P_{\phi}(k)P_{\phi}(|{\bm k} - {\bm q}|)\right],\\
 B_{\Phi}(k_1,k_2,k_{12}) & \equiv 
    2f_{nl}\,\left[ P_{\phi}(k_1)P_{\phi}(k_2) + 
    {\rm cyclic} \right] + \mathcal{O}(f_{nl}^3) \label{eqn:bkphi}
\end{align}
\citep{ssz04}.

Most of the complication in $f_{nl}$ models arises from the fact 
that we are almost always interested in spatially smoothed quantities.  
In particular, the quantity
 $\sigma S_3\equiv \langle \delta^3|R\rangle/\langle \delta^2|R\rangle^{3/2}$, 
will play an important role, because it represents the leading order 
contribution to the non-Gaussianity (note that it is proportional to 
$f_{nl}$).  Fortunately, smoothing is a linear operation, so the 
smoothed variables are just linear combinations of the unsmoothed ones.  
Hence, if $W(kR)$ denotes the Fourier transform of the smoothing 
window of scale $R$, then, to second order in $f_{nl}$, 
\begin{eqnarray}
 \langle \delta^2|R \rangle & = & \sigma^2(R) 
    = \frac{1}{(2\pi)^3}\int \frac{{\rm d} k}{k} 
       4\pi\, k^7 M^2(k)\,P_{\Phi}(k)\, W^2(kR), \qquad{\rm and}
   \label{eqn:variance} \\
 \langle \delta^3|R \rangle & = & 
   2 f_{nl} \,\frac{2}{(2\pi)^4}
               \int  \frac{{\rm d} k_1}{k_1} k_1^5 M(k_1) W(k_1R)
               \int  \frac{{\rm d} k_2}{k_2} k_2^5 M(k_2) W(k_2R)
               \int {\rm d}\mu_{12}\, k_{12}^2\, M(k_{12}) W(k_{12}R) 
                \frac{B_{\Phi}(k_1,k_2,k_{12})}{2f_{nl}} \label{eqn:skewness}
\end{eqnarray}
where $M(k) \equiv (3D(z)c^2)/(5\Omega_m H_0^2)\,T(k)$, 
$T(k)$ is the CDM transfer function and $D(z)$ is the linear growth function.  
In hierarchical models, $\sigma$ and $\sigma S_3$ are both 
monotonically decreasing functions of $R$; this will be important in 
what follows.  Appendix~\ref{sec:sigS3} provides a useful fitting 
formula for $\sigma\,S_3$, and shows that it is only a weak function 
of scale. 

\subsection{Edgeworth approximations for $p(\delta|R)$ and 
 $p(\delta,\Delta|r,R)$}
Because we are interested in small departures from Gaussianity, 
the Edgeworth expansion provides a convenient form for the 
distribution of $\delta$ smoothed on scale $R$:  
\begin{equation}
 p(\delta|R)\, {\rm d}\delta \approx \left[
 1 + \frac{\sigma(R)S_3(R)}{6}H_3\left(\frac{\delta}{\sigma(R)}\right)\right]\ 
  \frac{e^{-\delta^2/2\sigma^2(R)}}{\sqrt{2\pi}\sigma(R)}\,{\rm d}\delta
  = \left[1 + \frac{\sigma S_3}{6}H_3(\nu)\right]\,
    p_0(\delta|R)\, {\rm d}\delta,
\label{eqn:pdfdeltal}
\end{equation}
where $\sigma(R)$ is given by equation~(\ref{eqn:variance}), 
$\sigma S_3 \equiv \langle x^3\rangle/\langle x^2\rangle^{3/2} 
                 = 2f_{nl}\gamma^3/\sigma^3$,
and $H_3(\nu) \equiv \nu(\nu^2 - 3)$ with $\nu\equiv \delta/\sigma(R)$.  
Because $H_3$ changes sign at $\nu = \sqrt{3}$, the Edgeworth expansion
is not always positive, making it ill-suited for studying (at least
one of) the tails of the distribution.  The expansion becomes negative
when $(\sigma S_3/6) \nu (\nu^2-3)= -1$, meaning
$\nu (\nu^2-3) =  -200\, (0.03/\sigma S_3)$, and note that $\sigma S_3$
has the opposite sign to $f_{nl}$.  For $f_{nl}\approx 100$, we have
$\nu (\nu^2-3) = 200$, so there certainly are problems at $\nu>6$,
making the expansion suspect at slightly smaller values.
For $f_{nl}\approx -100$, there are problems at $\nu<-6$.
Because the Gaussian piece falls exponentially with $\nu^2$, the fact
that the expansion may not be accurate at large $\nu$ may not
matter -- but for larger $|f_{nl}|$ values, this limitation of the
Edgeworth approach should be borne in mind.

\cite{lamshethfnl} have used this expansion in the context of
modelling the one-point distribution of the evolved nonlinear
$\delta$, where $\sigma S_3$ is smaller, and large $\nu$ values are
indeed rare.  Here however, we will follow \cite{fnlverde}, and use
it in our model of halo abundances.  In this case, it is the large
$\nu$ tail which is of most interest, and this is precisely where
the Edgeworth expansion is most suspect.
However, note that, when modeling halos, one is most interested in
the regime where $\nu>0$, so, for $f_{nl}<0$, the Edgeworth
expansion is positive definite except for large $\sigma S_3$,
for which we know the Edgeworth expansion is not useful anyway.
In any case, Appendix~\ref{sec:sigS3} shows that large values of $\sigma S_3$ are 
not currently a concern.
%See \cite{fnlverde} and \cite{lamshethfnl} for previous work with the 
%Edgeworth expansion in the context of $f_{nl}$ models.

For reasons that will become clear shortly, we will also be 
interested in the value of the field when it is smoothed on two 
different scales.  For small departures from Gaussianity, the 
bivariate Edgeworth expansion should provide a good description.  
It is 
\begin{equation}
p(\mu,\nu) = \frac{1}{2\pi\sqrt{1-q^2}}
   \exp\left[-\frac{\mu^2-2q\mu\nu + \nu^2}{2(1-q^2)}\right]\,
   \left[1 + \frac{\lambda_{30}H_{30} + \lambda_{03}H_{03}}{6} + % 
         \frac{\lambda_{21}H_{21} + \lambda_{12}H_{12}}{2}\right]
\label{eqn:edgeworth2}
\end{equation}
\citep{kbj00}, where 
\begin{equation*}
 \mu  \equiv \frac{\delta}{\langle\delta^2\rangle^{1/2}}, \quad 
 \nu  \equiv \frac{\Delta}{\langle\Delta^2\rangle^{1/2}}, \quad 
  q   \equiv \langle \mu\,\nu\rangle, \quad
  \lambda_{mn} = \langle\mu^m\nu^n\rangle_{\, \rm c},
 \quad {\rm and}\quad
 H_{mn}(\mu,\nu,q) =  \frac{h_{mn}(\mu,\nu,q)}{(1-q^2)^2}, 
\end{equation*}
with
\begin{eqnarray*}
 h_{30}(\mu,\nu,q)  &=&  h_{03}(\nu,\mu,q) = \frac{(\mu - q\nu)^3}{1-q^2} %
                        - 3(\mu-q\nu), \quad \nonumber\\
 h_{21}(\mu,\nu,q)  &=&  h_{12}(\nu,\mu,q) 
                     =  2q(\mu-q\nu) - (\nu-q\mu) %
                        + \frac{(\nu-q\mu)(\mu-q\nu)^2}{1-q^2}.
\end{eqnarray*}
If our convention is that $\Delta$ is the field on the larger 
smoothing scale, then, to lowest order in $\lambda_{03} = \sigma S_3$,
\begin{equation}
p(\mu|\nu) = 
          \frac{1}{\sqrt{2\pi(1-q^2)}}  %
                 \exp\left[-\frac{(\mu-q\nu)^2}{2(1-q^2)}\right]\,
   \Biggl[1 + \frac{\lambda_{30}H_{30} + \lambda_{03}H_{03}}{6} 
            + \frac{\lambda_{21}H_{21} + \lambda_{12}H_{12}}{2} 
            - \frac{\lambda_{03}H_3(\nu)}{6}\Biggr].
\label{eqn:bivarcon}
\end{equation}
In what follows, we will set 
 $q^2 = \langle\Delta^2\rangle/\langle\delta^2\rangle$ (as it is for 
a Gaussian field smoothed with a tophat in $k$-space; this is standard 
for the excursion set approach).

\section{Halo abundances}\label{section:mf}
Recall that $\sigma$ is a monotonically decreasing function of 
smoothing scale $R$.  In the initial conditions, where fluctuations 
are negligible, the smoothing scale $R$ contains mass 
$m=\bar\rho\,4\pi R^3/3$ almost surely, so $\sigma$, $R$ and $m$ 
are equivalent variables.  
Now, let $dn/dm$ denote the comoving number density of halos of mass $m$, 
and let $\bar\rho$ denote the comoving density of the background.  
If $f(m)$ denotes the mass fraction in halos of mass $m$, then 
\begin{equation}
 F\left(<\sigma\left(R=(3m/4\pi\bar\rho)^{1/3}\right)\right)
   = F(>m) = \int_m^\infty dm\,f(m) 
   = \int_m^\infty \frac{dm}{\bar\rho}\,\frac{dn(m)}{d\ln m}.
\end{equation}
An estimate of $F$, then, is an estimate of $dn/d\ln m$.  
The following sections describe the excursion set estimate of $F$ 
for Gaussian initial conditions, and how this estimate can be 
extended to the $f_{nl}\ne 0$ models.  

\subsection{Excursion set approach}
The excursion set approach \citep{bcek91,laceycole93,rks98} 
relates the number of haloes of mass $m$ to the first crossing of 
a suitably chosen barrier $b(\sigma)$ by a suitably chosen set of 
walks.  The simplest implementations of this approach consider the 
first crossing of $b(\sigma)$ by an ensemble of uncorrelated random 
walks with uncorrelated steps.  While neglecting both types of 
correlations is far from ideal, previous work shows that this 
allows one to write down simple analytic expressions for the 
first crossing distribution and how this distribution is related 
to halo abundances, which together provide reasonably accurate 
descriptions of halo abundances as well as their formation histories.  

The barrier shape $b(\sigma)$ is set by the physics of collapse.  
The spherical collapse model has a barrier of constant height 
$b(\sigma) = \delta_{\rm sc}$, whereas barriers of the form 
\begin{equation}
 b(\sigma) = \sqrt{a}\delta_c[1 + \beta\,(\sigma/\sqrt{a}\delta_c)^{2\gamma}], 
 \label{eqn:dec}
\end{equation}
with $\beta=0.4$ and $\gamma=0.6$ may be related to models in which 
halos from from a triaxial collapse \citep{smt01,st02}.  The physics 
of collapse has $a=1$, but setting $a=0.7$ results in a predicted 
$dn/d\ln m$ which is in much better agreement with the abundance 
observed in simulations.  \citep[See][for further discussion of why 
$a\ne 1$]{smt01}.

\subsection{The Gaussian case: Spherical collapse}\label{betterPS}
Let $p(\delta,s)$ denote the probability that a randomly placed cell 
in the initial distribution has overdensity $\delta$ when the 
smoothing scale is such that $\langle\delta^2\rangle = \sigma^2 \equiv s$.  
Classify all cells by the largest scale $S\le s$ on which they had 
overdensity greater than $\delta_c$.  Then, provided $\delta\ge\delta_c$, 
\begin{equation}
 p(\delta,s) = \int_0^s dS\,f(S,\delta_c)\,p(\delta,s|\delta_c,S,{\rm first}),
\end{equation}
where $f$ is the fraction of cells for which $S$ was the largest 
smoothing scale on which the overdensity was greater than $\delta_c$, 
and $p(\delta,s|\delta_c,S,{\rm first})$ is the probability that the 
overdensity on scale $s$ is $\delta$ given that $S$ was the largest 
scale on which the overdensity exceeded $\delta_c$.  If one views a 
plot of $\delta$ versus smoothing scale as something which resembles 
a random walk, with large smoothing scales to the left, then $S$ is 
the first `time' that the walk crosses $\delta_c$:  hence the word 
`first' in the expression above.  Therefore
\begin{equation}
 P(\delta_c,s) \equiv \int_{\delta_c}^\infty d\delta\,p(\delta,s) 
  = \int_0^s dS\,f(S,\delta_c)\,
   \int_{\delta_c}^\infty d\delta\, p(\delta,s|\delta_c,S,{\rm first}) 
  = \int_0^s dS\,f(S,\delta_c)\,P(\delta_c,s|\delta_c,S,{\rm first}) .
  \label{pf}
\end{equation}

In what follows, we will use the subscript `0' to denote quantities 
associated with Gaussian initial conditions (for which $f_{nl}=0$).  
For a Gaussian field smoothed with a tophat in $k$-space, 
\begin{equation}
 p_0(\delta_1,s|\delta_2,S,{\rm first}) = p_0(\delta_1,s | \delta_2,S) 
  = p_0(\delta_1-\delta_2,s-S), \label{eqn:gauconP}
\end{equation}
so $P_0(\delta_c,s|\delta_c,S,{\rm first}) = P_0(0,s-S) = 1/2$, and 
equation~(\ref{pf}) implies 
\begin{equation}
 P_0(\delta_c,s) \equiv \int_{\delta_c}^\infty d\delta\,p_0(\delta,s) 
 = \int_0^s dS\,\frac{f_0(S,\delta_c)}{2}
 \equiv \frac{F_0(<s)}{2} = \frac{F_0(>m)}{2}.
 \label{eqn:ps2}
\end{equation}
Differentiating both sides with respect to $s$ shows that $f_0$ is 
simply related to $p_0$:
\begin{equation}
 \frac{\partial P_0}{\partial s} = 
 -p_0\left(\frac{\delta_c}{\sqrt{s}}\right)
 \,\frac{\partial(\delta_c/\sqrt{s})}{\partial s}
  = \frac{\delta_c}{2s^{3/2}}\frac{{\rm e}^{-\delta_c^2/2s}}{\sqrt{2\pi}}
  = \frac{f_0(s,\delta_c)}{2}.
 \label{ps3}
\end{equation}

\subsection{The Gaussian case:  Ellipsoidal collapse}\label{betterST}
If $\delta_c$ depends on $s$, as in some parametrizations of 
triaxial collapse, then this simplicity is lost.  
In particular, for the barrier given in equation~\eqref{eqn:dec}, 
\cite{st02} show that 
\begin{equation}
 sf_0(s,b) \approx \frac{b(0)}{\sqrt{2\pi s}}\exp\left[-\frac{b^2}{2s}\right]
  \left[1 + 0.067\,\frac{s^{\gamma}}{(a\delta_c^2)^{\gamma}}\right].
 \label{eqn:st}
\end{equation}
To see where this comes from, note that the analogue of 
equation~\eqref{eqn:ps2} becomes 
\begin{equation}
 \frac{\partial P_0}{\partial s} = 
 -\frac{{\rm e}^{-b^2/2s}}{\sqrt{2\pi s}}\,
    \left[\frac{\partial b}{\partial s} - \frac{b}{2s}\right]
  = \frac{f_0(s,b)}{2} - \int_0^s dS\,f_0(S,B)\,
   \frac{{\rm e}^{-(b-B)^2/2(s-S)}}{\sqrt{2\pi(s-S)}}
        \left[\frac{\partial b}{\partial s} -  \frac{(b-B)}{2(s -S)}\right].
\end{equation}
However, the first term in square brackets on the right hand side is 
independent of $S$, so this term integrates to
 $p(b,s)\,\partial b/\partial s$.  
Since there is a similar term on the left hand side, this leaves 
\begin{equation}
 \frac{b}{s}\,\frac{{\rm e}^{-b^2/2s}}{\sqrt{2\pi s}}\,
 = f_0(s,b) + \int_0^s dS\,f_0(S,B)\,
   \frac{{\rm e}^{-(b-B)^2/2(s-S)}}{\sqrt{2\pi(s-S)}} \frac{(b-B)}{(s -S)}.
\end{equation}
where we have multiplied both sides by a factor of 2.  Writing 
$B(S)$ as a Taylor series around $b(s)$ 
implies 
\begin{equation}
 \frac{b}{s}\,\frac{{\rm e}^{-b^2/2s}}{\sqrt{2\pi s}}\,
 = f_0(s,b) +  \sum_{i=1}^\infty \frac{\partial^i b}{\partial s^i} 
    \int_0^s dS\,f_0(S,B)\,
   \frac{{\rm e}^{-(b-B)^2/2(s-S)}}{\sqrt{2\pi(s-S)}} \frac{(S-s)^{i-1}}{i!}.
\end{equation}
so 
\begin{equation}
 sf_0(s,b) = \left[b - s\frac{\partial b}{\partial s}\right]\,
  \frac{{\rm e}^{-b^2/2s}}{\sqrt{2\pi s}}\,
  - \sum_{i=2}^\infty \frac{s^i}{i!}\,
               \frac{\partial^i b}{\partial s^i} 
    \int_0^s dS\,f_0(S,B)\,
   \frac{{\rm e}^{-(b-B)^2/2(s-S)}}{\sqrt{2\pi(s-S)}}\, (S/s - 1)^{i-1}.
\end{equation}
Equation~(\ref{eqn:st}), the approximation of \cite{st02}, 
corresponds to ignoring all $S/s$ terms and then keeping only the 
first few terms in the series.

\subsection{Dependence on $f_{nl}$}
We now extend the analysis above to the case in which $f_{nl}\ne 0$.
Our primary assumption is that the expressions above remain valid 
if $p$ on the left hand side of equation~(\ref{pf}) is given by the 
Edgeworth expansion, and the conditional distribution on the right 
hand side is given by the bivariate Edgeworth expansion.  This is 
not quite right, since the steps in the walk are now correlated, 
and we are assuming that the statistics of walking from $(B,S)$ to 
$(b,s)$ do not depend on the fact that the walk did not cross $B$ 
before $S$.  In effect, this means that we assume that the extra 
constraint that $\delta'<\delta_c$ for all $S'<S$ makes no difference, 
so the only difference from the Gaussian calculation is that all PDFs 
are now replaced by the appropriate Edgeworth approximation.  

To see what this implies, it is convenient to first define
\begin{equation}
\mathcal{G}_{mn} = \int^{\infty}_0 {\rm d}\delta\, p_0(\delta+b,s|B,S)
   h_{mn}\left(\frac{\delta+b}{\sqrt{s}}, \frac{B}{\sqrt{S}},\sqrt{\frac{S}{s}} \right),
\end{equation}
where
\begin{align}
\mathcal{G}_{30} & = -\frac{\sqrt{s-S}}{\sqrt{s}}
  \left[1 -\frac{(b-B)^2}{s-S}\right]p_0\left(\frac{b-B}{\sqrt{s-S}}\right)
 \nonumber \\
\mathcal{G}_{03} & = \frac{(s-S)^2}{s^2}
    H_3(B/\sqrt{S}) P_0\left(\frac{b-B}{\sqrt{(s-S)}}\right) \nonumber \\ 
& \quad - \frac{1}{s^2\sqrt{S}\sqrt{s-S}}
   \left\{ S(s-S)(2S-3s) + [S^2(b^2+bB+B^2)-3sS(bB+B^2)+3B^2s^2]\right\}
   p_0\left(\frac{b-B}{\sqrt{s-S}}\right) \nonumber \\
\mathcal{G}_{21} &= \frac{\sqrt{S}\sqrt{s-S}}{s}
       \left[1 - \frac{(b-B)^2}{s-S} + \frac{B(b-B)}{S}\right]
    p_0\left(\frac{b-B}{\sqrt{s-S}}\right) \nonumber \\
\mathcal{G}_{12} &=\frac{-sS(s-S) + (bS - Bs)^2}{Ss^{3/2}\sqrt{s-S}}
   p_0\left(\frac{b-B}{\sqrt{s-S}}\right).
\end{align}
In addition, define 
\begin{equation}
\mathcal{G}_3\equiv \int^{\infty}_0 {\rm d}\delta\, 
     p_0(\delta+b,s|B,S)H_3(B/\sqrt{S}) 
 = H_3(B/\sqrt{S}) P_0\left(\frac{b-B}{\sqrt{s-S}}\right).
\end{equation}
If we ignore the scale dependence of $\sigma S_3$ (Figure~\ref{fig:sigs3} 
shows that it is weak), then the derivative of equation~\eqref{pf} 
w.r.t. $s$ is
\begin{equation}
\frac{\partial P_0(b/\sqrt{s})}{\partial s}
 \left[1 + \frac{\sigma S_3}{6}H_3\left(\frac{b}{\sqrt{s}}\right)\right]
 = \frac{f(s,b)}{2} 
 + \int^s_0 {\rm d}S\, f(S,B)\frac{\partial}{\partial s}
                         P_0\left(\frac{b-B}{\sqrt{s-S}}\right) 
 + \frac{\sigma S_3}{6}\int^s_0 {\rm d}S\, f(S,B)
            \frac{\partial}{\partial s}\left[ \mathcal{E}(s,S)  
                p_0\left(\frac{b-B}{\sqrt{s-S}}\right)\right],
 \label{p0fsfnl}
\end{equation}
where 
\begin{equation}
\mathcal{E}(s,S) = \frac{-4s^{3/2} + 6s\sqrt{S} - 2S^{3/2}}{(s-S)^{3/2}}
+ \frac{3B\sqrt{s-S}}{S(\sqrt{s}+\sqrt{S})^2}\left(B\sqrt{s} +b\sqrt{S}\right)
+ \frac{(b-B)^2\sqrt{s-S}}{(\sqrt{s} + \sqrt{S})^3} .
\end{equation}
Notice that, in contrast to the case in which $f_{nl}=0$, the 
final term on the right hand side of equation~\eqref{p0fsfnl} is 
non-trivial because it is now $s$-dependent.  

\begin{figure}
\centering
\includegraphics[width=0.7\textwidth]{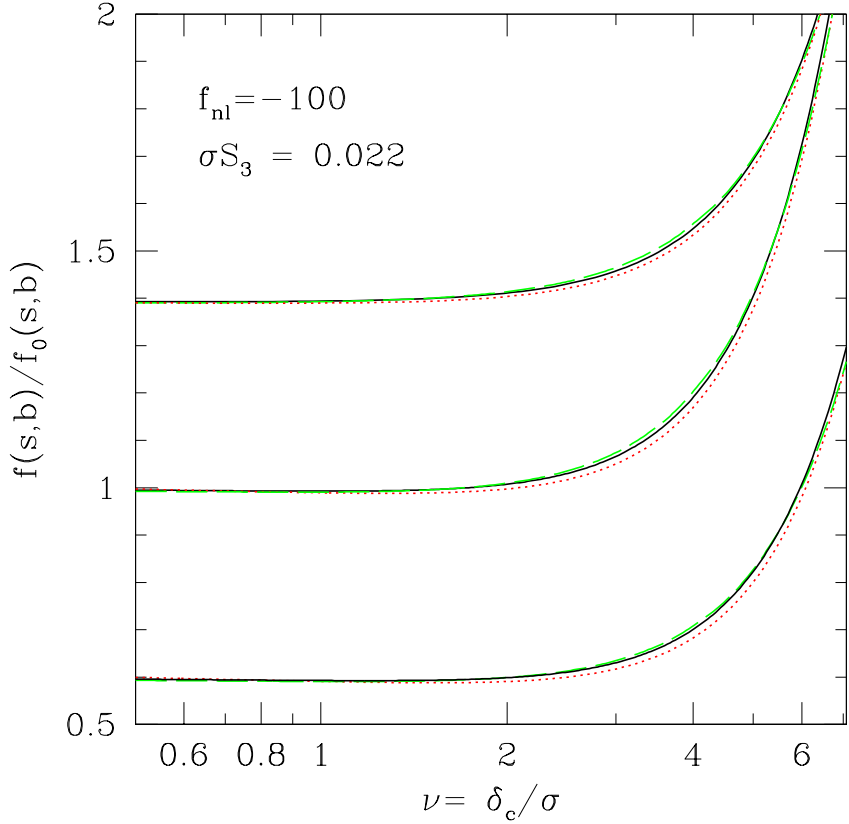}
\caption{Ratio of the first crossing probability for $f_{nl}=-100$ to 
         that when $f_{nl}=0$. 
         Solid (black) curves set $\sigma S_3 = 0.022$ and  
         $\mathcal{G}(s,b)=0$; 
         dotted (red) curves include $\mathcal{G}(s,b)\ne 0$ 
         but keep $\sigma S_3$ fixed, and 
         the dashed (green) curves include the scale dependence of 
         $\sigma S_3$ but set $\mathcal{G}(s,b)=0$.
         The top set of curves (offset upwards by 0.4) show results 
         for a moving barrier (equation~\ref{eqn:dec}); 
         middle curves use a constant barrier with $\delta_c = 1.66$ 
         and lowest curves (offset downwards by 0.4) use a constant 
         barrier of height $\sqrt{0.7}\delta_c$.
         }
\label{fig:gyh3ALL}
\end{figure}

Substituting the $f_{nl}=0$ solution for the pieces with subscript zero 
yields 
\begin{equation}
\begin{split}
\frac{f_0(s,b)}{2} 
\left[1 + \frac{\sigma S_3}{6}H_3\left(\frac{b}{\sqrt{s}}\right)\right]
& = \frac{f(s,b)}{2}\left\{ 1 + 2\int^s_0{\rm d}S\,
        \frac{\partial}{\partial s} P_0\left(\frac{b-B}{\sqrt{s-S}}\right)
\frac{f(S,B) - f_0(S,B)[1 + (\sigma S_3/6) H_3(b/\sqrt{s})]}{f(s,b)} \right.\\
& \quad \quad \left. \qquad\qquad + 2 \frac{\sigma S_3}{6}\int^s_0{\rm d}S\,
  \frac{f(S,B)}{f(s,b)}
    \frac{\partial}{\partial s}\left[ \mathcal{E}(s,S) 
       p_0\left(\frac{b-B}{\sqrt{s-S}}\right)  \right]
\right\}.\\
\end{split}
\label{eqn:esfnl}
\end{equation}
This is an integral equation for $f(s,b)$ which is valid when $f_{nl}\ne 0$.  
Clearly, the zeroth-order solution is simply the $f_{nl}=0$ solution 
(the Gaussian case) times a correction term which depends on 
$\sigma S_3$ and on the barrier shape $b(s)$.  
We can include the next-to-leading order contribution as
\begin{equation}
f(s,b) = f^{(0)}(s,b)\left( 1 + \frac{f^{(1)}}{f^{(0)}}\right) = 
 f_0(s,b)\left[1 + \frac{\sigma S_3}{6}
                     H_3\left(\frac{b}{\sqrt{s}}\right)\right]
          \left[1  - \frac{\sigma S_3}{6}\,\mathcal{G}(s,b)\right] ,
\end{equation}
where $f_0$ is the first crossing probability associated with 
uncorrelated steps when $f_{nl}=0$ (approximated by 
equation~\ref{eqn:st}), 
\begin{equation}
\mathcal{G}(s,b)  = 2\int^s_0{\rm d}S\,
   \frac{f_0(S,B)}{f_0(s,b)}  \left\{
   \frac{\partial}{\partial s}\left[ \mathcal{E}(s,S) 
       p_0\left(\frac{b-B}{\sqrt{s-S}}\right)  \right]
  + \frac{\partial}{\partial s} P_0\left(\frac{b-B}{\sqrt{s-S}}\right)
   \left[H_3(B/\sqrt{S}) - H_3(b/\sqrt{s})\right]
   \right\},
\label{eqn:ggy}
\end{equation}
where only terms to first order in $\sigma S_3$ have been kept, and 
\begin{equation}
\begin{split}
\frac{\partial \mathcal{E}}{\partial s} = &
    \frac{-3\sqrt{S}}{\sqrt{s-S}(\sqrt{s}+\sqrt{S})^2}
  + \frac{3B}{S\sqrt{s-S}(\sqrt{s}+ \sqrt{S})}
      \left[\frac{3B(\sqrt{s}-\sqrt{S})}{2\sqrt{s}} 
        + (B+b)\sqrt{\frac{S}{s}} 
        + \sqrt{S}\frac{\partial b}{\partial s}(\sqrt{s} - \sqrt{S})\right]\\
& -\frac{9B}{2S\sqrt{s-S}(\sqrt{s}+\sqrt{S})^2}\left(B\sqrt{s}+b\sqrt{S}\right)
+ \frac{b-B}{\sqrt{s-S}(\sqrt{s}+\sqrt{S})^3}
    \left[2(s-S)\frac{\partial b}{\partial s} + \frac{b-B}{2}\right]
 - \frac{3}{2}\frac{(b-B)^2\sqrt{s-S}}{\sqrt{s}(\sqrt{s}+\sqrt{S})^4}.
\end{split}
\end{equation}
Thus, 
\begin{equation}
 f(s,b) \approx f_0(s,b) 
 \left[1 + \frac{\sigma S_3}{6}\,H_3\left(\frac{b}{\sqrt{s}}\right)
         - \frac{\sigma S_3}{6}\,\mathcal{G}(s,b)\right].
 \label{fs3gs}
\end{equation}
The analysis simplifies somewhat for a constant barrier -- see 
Appendix~\ref{constant}.  Previous analyses of the case in which the 
barrier is a constant have either explicitly ignored the final term in 
the square brackets above \citep{mvj00}, or have missed the fact that 
it is there entirely \citep{fnlverde}.

Figure~\ref{fig:gyh3ALL} shows the effect of $\mathcal{G}(s,b)$ on 
the ratio of the first crossing distribution when $f_{nl}=-100$ to 
that when $f_{nl}=0$.  This is done for three different barriers:  
the top curves (offset upwards by 0.4) show results for the 
moving barrier of equation~\eqref{eqn:dec}, 
the middle are for a constant barrier of height $b = \delta_c$, 
and the lowest (offset downwards by 0.4) for $b =\sqrt{0.7}\delta_c$.  
The middle curves are for the barrier shape that has been 
previously studied \citep{mvj00, fnlverde}; differences 
between these and the bottom curves are due to the barrier height.  
Differences between the top and bottom sets of curves are due to 
the $s$ dependence of the barrier.  

In each case, the black (solid) curves show the result of neglecting 
the factor of $\mathcal{G}(s)$ and setting $\sigma S_3 = 0.022$ (i.e. 
we ignore the fact that it depends weakly on scale).   
The red (dotted) curves include the fact that $\mathcal{G}(s,b)\ne 0$ 
(or the corresponding $G(s)$ (equation~\ref{eqn:gy}) for the constant 
barrier) with the same (fixed) value of $\sigma S_3$.  
The differences between these two sets of curves are small for all 
three barriers, indicating that the contribution from the factor 
$\mathcal{G}$ is small. 
The effect of again ignoring $\mathcal{G}$, but now including the 
scale dependence of $\sigma S_3$ (using equation~\ref{eqn:sigs3}), 
but setting $\partial\sigma S_3/\partial s \approx 0$ is shown 
by the green (dashed) curves.

\begin{figure}
\centering
\includegraphics[width=0.7\textwidth]{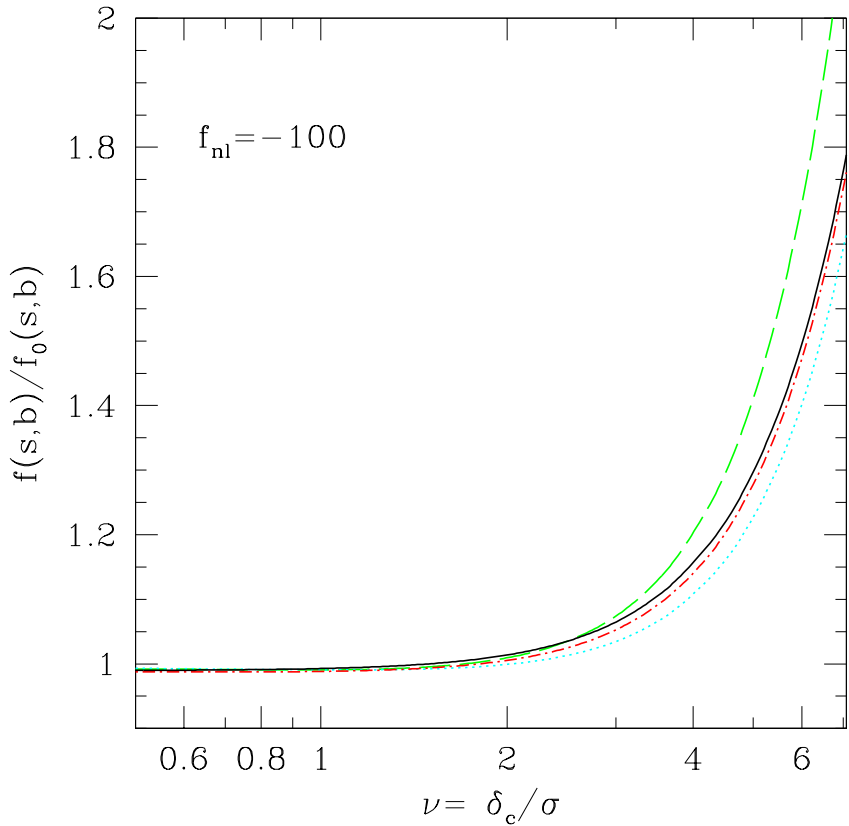}
\caption[The effect of using different barrier shapes on $f(s,b)$]{
         Ratios of the first crossing probability for $f_{nl}= -100$  
         to that when $f_{nl}=0$. Black (solid), green (dashed), and 
         cyan (dotted) curves plot the ratios for moving barrier $b(s)$, 
         $b=\delta_c$ and $b = \sqrt{0.7}\delta_c$ when $\mathcal{G}(s)$ 
         is set to $0$. The red (dot-dashed) curve shows 
         equation~\eqref{fs3gs}.
         }
\label{fig:gyh3ss3}
\end{figure}
Figure~\ref{fig:gyh3ss3} shows the ratios of the first crossing 
distribution for different barrier shapes. The scale dependence of 
$\sigma S_3$ is included and computed by the approximation formula 
(equation~\ref{eqn:sigs3}). Black (solid), green (dashed), and 
cyan (dotted) curves plot the ratios for moving barrier $b(s)$, 
$b=\delta_c$, and $b=\sqrt{0.7}\delta_c$ respectively (all without 
$\mathcal{G}$). The red (dot-dashed) curve shows 
the term in square brackets in equation~(\ref{fs3gs}):  
this includes the scale dependence in the factor $\sigma S_3$ 
but it sets $\partial \ln\sigma S_3/\partial s \approx 0$, and it 
ignores the fact that this scale dependence will also modify the 
$\mathcal{G}(s,b)$ term. 
The inclusion of the factor $\mathcal{G}(s,b)$ has a small effect 
compared to  using  different barrier shapes. 
The three barrier shapes result in slightly different predictions 
for how $f_{nl}$ modifies halo abundances.  Therefore, comparisons 
with measurements in numerical simulations may indicate which barrier 
shapes better describe halo formation.

Figures~\ref{fig:gyh3ALL} and \ref{fig:gyh3ss3} show that the effect 
of $\mathcal{G}(s,b)$ is small compared to the effects of including 
the scale dependence of $\sigma S_3$, and the effect of using 
different barrier shapes.  Therefore, it is a good approximation 
to set 
\begin{equation}
 f(s,b) \approx f_0(s,b) \,
 \left[1 + \frac{\sigma S_3}{6}H_3\left(\frac{b}{\sqrt{s}}\right)\right].
\end{equation}
When the barrier is constant, $b=\delta_c$, then this reduces to 
the expression presented by \cite{fnlverde}.  Our analysis shows 
that there is an additional correction factor which their derivation 
missed (our factor of $(\sigma S_3/6)\,G$), but that this happens 
to be small.  On the other hand, they include a term which comes from 
the scale dependence of $\partial (\sigma S_3/6)/\partial\ln s$ which 
we are ignoring (we have checked that it is small).

%%%%%%%%%%%%%%%%%%%%%%%%%%%%%%%%%%%%%%%%%%%%%%%%%%%%%%%%%%%%%%%%%%%%%%%
\section{Discussion}
We showed how the excursion set approach may be extended to model 
halo abundances when the initial conditions were non-Gaussian.  
In this approach, the estimate of halo abundances is related to 
the first crossing distribution of a suitably chosen barrier by 
a suitably chosen set of walks.  The physics of collapse sets the 
barrier shape, and the statistics of the initial fluctuation field 
sets the properties of the ensemble of walks:  e.g., how steps in a 
given walk are correlated, and whether the walks are independent 
of one-another.  Our analysis assumes that the steps in a walk are 
uncorrelated, and the appropriate ensemble contains all possible 
independent walks.  The first assumption may be well-motivated only 
for Gaussian random fields -- we argue that it may remain a useful 
approximation for weakly non-Gaussian fields.  The second assumption 
ignores the fact that averaging over the full ensemble of uncorrelated 
walks is only an approximation to the more physically appropriate 
ensemble described in \cite{smt01}.  Previous work suggests that 
this allows one to write down simple analytic expressions which are 
reasonably accurate.  

Our analysis was done in two steps -- the first showed how the 
calculation depends on the physics of collapse:  spherical and 
triaxial collapse models are associated with `constant' and `moving' 
barriers (Section~\ref{section:mf}).
For Gaussian initial conditions, our results clarify the nature of 
approximations made in previous studies of the moving barrier model
(see Section \ref{betterST}).  

For non-Gaussian initial conditions, our analysis assumes that the 
correlated nature of the steps (in non-Gaussian models) changes the 
step-size distribution in a calculable way, but that steps are 
otherwise independent.  For weak non-Gaussianity, we approximate the 
change to the step-sizes by using the Edgeworth expansion  
(equations~\ref{eqn:pdfdeltal} and~\ref{eqn:edgeworth2})
This is not strictly correct -- it is an approximation which may be 
accurate for weakly non-Gaussian fields.  
(In this context, our analysis of the constant barrier model showed 
that previous work on this problem, which made the same assumptions 
as we do, had missed some terms.  However, these turn out to be small.)  
As we were completing this work, \citet{mr09c} presented a very 
different analysis of the constant barrier problem which yields 
consistent results, suggesting that our neglect of the additional 
correlations between steps associated with $f_{nl}\ne 0$ is reasonable.  

Ours is the first analysis of the moving barrier problem for 
non-Gaussian models:  the moving barrier yields quantitatively 
different predictions for halo abundances than does the constant 
barrier model, at a level that current simulations should be able 
to detect.  This is explored further in \cite{LSDfnl}.  So we hope 
that our results will benefit problems which use halo abundances to 
constrain the nature of the initial fluctuation field.  They also 
provide a key ingredient to the Halo Model interpretations of how 
galaxies cluster \citep{haloreview}.

%\bibliography{LamSheth}

%%%%%%%%%%%%%%%%%%%%%%%%%%%%%%%%%%%%%%%%%%%%%%%%%%%%%%%%%%%%%%%%%%%%%%%
\appendix

\begin{figure}
\centering
 \includegraphics[width=0.8\linewidth]{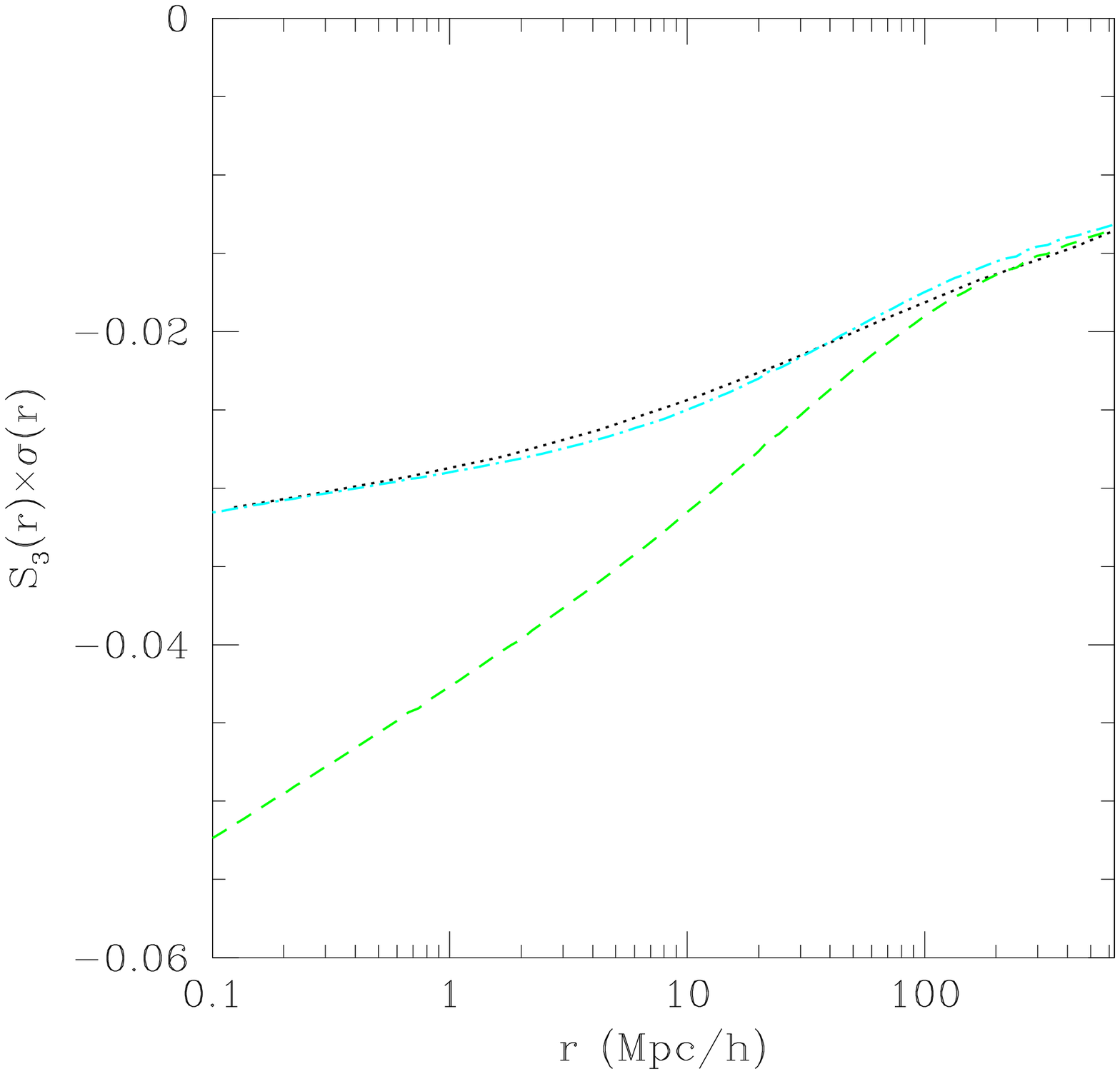}
 \caption{The quantity $\sigma S_3(r)$ for $f_{nl}=100$.  
          Dotted (black) curve is the numerical result, 
          dot-dashed (cyan) curve is equation~(\ref{eqn:sigs3}), and 
          dashed (green) curve is the approximation from 
          \citet{ssz04}, which is only expected to be valid on large 
          scales.}
 \label{fig:sigs3}
\end{figure}

\section{Approximation formula for $\sigma S_3$}\label{sec:sigS3}
The quantity $\sigma S_3$ measures the strength of non-Gaussianity in 
the smoothed field.  Unfortunately, it requires the computation of 
several numerical integrals.  We have found that the following provides 
a good approximation: 
\begin{equation}
\sigma S_3 \approx 12\,f_{nl} 
            \frac{\langle \delta_R \phi_R \rangle}{\sigma}\,
            \left(1 + \frac{1}{6} %
            \frac{{\rm d}\,\ln\langle \delta_R \phi_R \rangle}%
            {{\rm d}\,\ln r}\right)\left[%
            1 - \left(\frac{\sigma}{\sigma(r_0)}\right)^{m_1}\right]
       + \sigma S_3(r_0) 
   \frac{\langle \delta_R \phi_R \rangle/\langle \delta_R \phi_R \rangle(r_0)}%
     {\sigma/\sigma(r_0)}\left[\frac{\sigma}{\sigma(r_0)}\right]^{m_2}, 
\label{eqn:sigs3}
\end{equation}
where the first term on the right hand side is the large scale approximation 
given in \citet{ssz04}.  If we set $r_0 = 0.1 {\rm Mpc}/h$, $m_1=0.5$, 
$m_2 = 0.7$, then $\sigma S_3(r_0) = -0.0316$, and 
equation~(\ref{eqn:sigs3}) is rather accurate 
(see Figure~\ref{fig:sigs3}).

%%%%%%%%%%%%%%%%%%%%%%%%%%%%%%%%%%%%%%%%%%%%%%%%%%%%%%%%%%%%%%%%%%%%%

\section{Halo abundances in the constant barrier model when $f_{nl}\ne 0$}\label{constant}
The main text considered the general case of a moving barrier.  
When the barrier is constant then it is convenient to define 
\begin{equation}
G_{mn} \equiv 
 \int_0^{\infty} {\rm d}\delta\, p_0(\delta+\delta_c,s; \delta_c,S)
  h_{mn}\left( \frac{\delta+\delta_c}{\sqrt{s}}, 
                \frac{\delta_c}{\sqrt{S}},\sqrt{\frac{S}{s}}\right),
\end{equation}
making 
\begin{align}
G_{30} & = -\frac{\sqrt{s-S}}{\sqrt{2\pi}\sqrt{s}}, \qquad 
G_{03}  =  \frac{(\delta_c^3-3S\delta_c)(s-S)^2}{2s^2S^{3/2}}
   - \frac{\sqrt{s-S}[S(2S - 3s)+ 3\delta_c^2(s-S)]}{\sqrt{2\pi}s^2\sqrt{S}},
                                                    \nonumber \\
G_{21} & = \frac{\sqrt{S}\sqrt{s-S}}{\sqrt{2\pi}s}, \qquad {\rm and}\qquad
G_{12}  = \frac{-sS(s-S) + \delta_c^2(s-S)^2}{\sqrt{2\pi}Ss^{3/2}\sqrt{s-S}}.
\end{align}
In addition, 
\begin{equation}
 G_3 \equiv \int_0^{\infty} {\rm d}\delta\, p_0(\delta+\delta_c,s; \delta_c,S) 
     H_3(\delta_c/\sqrt{S}) =  \frac{(\delta_c^3-3S\delta_c)}{2S^{3/2}},
\end{equation}
so 
\begin{equation}
\int^{\infty}_{0} p(\delta + \delta_c,s|\delta_c,S) = 
  \frac{1}{2} + \frac{\lambda_{30}}{6}\frac{G_{30}s^2}{(s-S)^2}
    + \frac{\lambda_{03}}{6}
    \left[ \frac{\sqrt{S}(3s-2S)}{\sqrt{2\pi}(s-S)^{3/2}} - 
           \frac{3\delta_c^2}{\sqrt{2\pi}\sqrt{S}\sqrt{s-S}}\right] 
 + \frac{\lambda_{21}}{2}\frac{G_{21}s^2}{(s-S)^2} + 
   \frac{\lambda_{12}}{2}\frac{G_{12}s^2}{(s-S)^2}.
\label{eqn:intgeneral}
\end{equation}
If the scale dependence of $\sigma S_3$ can be ignored, 
then $\lambda_{mn} = \sigma S_3$, so, to first order in $\sigma S_3$, 
\begin{equation}
\int^{\infty}_{0} p(\delta + \delta_c,s|\delta_c,S) = 
\frac{1}{2} + \frac{1}{\sqrt{2\pi}}\frac{\sigma S_3}{6} E(s,S),
 \qquad {\rm where}\quad 
E(s,S) = \frac{2(-2s^{3/2} + 3s\sqrt{S}- S^{3/2})}{(s-S)^{3/2}}
             + \frac{3\delta_c^2(\sqrt{s}-\sqrt{S})}{S\sqrt{s-S}}.
\end{equation}
Note that $E(s,S)=0$ when $s=S$, so when $\sigma S_3$ is a constant, 
then 
\begin{equation}
\frac{\partial}{\partial s}\int_{\delta_c}^\infty d\delta\,p(\delta,s) 
 = \frac{f(s,\delta_c)}{2} + \frac{1}{\sqrt{2\pi}}
   \frac{\sigma S_3}{6} \int^{s}_0{\rm d}S\, f(S,\delta_c)
                                          \, \frac{\partial E}{\partial s},
   \quad {\rm where} \quad
   \frac{\partial E}{\partial s}
  = \frac{3}{2s}\sqrt{\frac{S}{s}} 
    \frac{(\delta_c^2/S) (1 +\sqrt{S/s}) - 2}{\sqrt{(1-S/s)}(1+\sqrt{S/s})^2}.
\label{eqn:haloF}
\end{equation}
The main text assumes that the second term on the right-hand side 
is negligible compared to the first.  In the approximation where 
$\sigma S_3$ is constant, this makes
$f(s,\delta_c) = f_0(s,\delta_c)\,[1 + (\sigma S_3/6)\,H_3(\delta_c/\sigma)]$. 
To see if this is accurate, one can substitute this expression for 
$f(s,\delta_c)$ into the integral, and check that this contribution 
really is negligible compared to the first term, $f(S,\delta_c)/2$. 
Namely, write 
\begin{equation}
\frac{\partial}{\partial \ln s}\int_{\delta_c}^\infty d\delta\,p(\delta,s) 
 = \frac{sf(s,\delta_c)}{2}\left[1 + \frac{2}{\sqrt{2\pi}}
   \frac{\sigma S_3}{6} \int^{s}_0 \frac{{\rm d}S}{S}\, 
            \frac{Sf(S,\delta_c)}{sf(s,\delta_c)}
                        \, \frac{\partial E}{\partial \ln s}\right],
\end{equation}
then note that, to leading order in $\sigma S_3$, we may approximate 
$f(S,\delta_c)\approx f_0(S,\delta_c)$, so the second term in 
square brackets is 
\begin{equation}
 G(s) = \frac{2}{\sqrt{2\pi}}\frac{\sigma S_3}{6}
   \int^{s}_0 \frac{{\rm d}S}{S}\, \frac{Sf_0(S,\delta_c)}{sf_0(s,\delta_c)}
    \frac{3}{2}\sqrt{\frac{S}{s}} 
    \frac{(\delta_c^2/S) (1 +\sqrt{S/s}) - 2}{\sqrt{(1-S/s)}(1+\sqrt{S/s})^2}
      = \frac{\sigma S_3}{2}\int_0^{1} \frac{{\rm d}x}{x}\,
        \frac{{\rm e}^{-(\delta_c^2/2s)(1/x-1)}}
                   {\sqrt{2\pi}}
    \frac{(\delta_c^2/s) (1 +\sqrt{x}) - 2x}{x\sqrt{1-x}(1+\sqrt{x})^2}.
\label{eqn:gy}
\end{equation}

\label{lastpage}
\end{document}